\documentclass{acm_proc_article-sp}

\usepackage{ifthen,hyperref} 

\newcommand{\draft}{false}

\ifthenelse{\equal{\draft}{true}}{

  \usepackage{longtable,amsthm,tabularx, longtable, booktabs,
    setspace,varioref, ifthen,natbib,amssymb,verbatim,dcolumn,
    graphicx,lscape,setspace,epsfig, epstopdf,
    amsmath,color,subfigure,threeparttable}
    
\oddsidemargin  0.0in
\evensidemargin 0.0in
\textwidth      7.0in
\headheight     0.0in
\topmargin      0.0in
\textheight=9.0in

\doublespacing 

\usepackage{lineno}
\linenumbers*[1] 
}{}

\usepackage{Sweave}
\begin{document} 
\title{The Condition of the Turking Class: \\ Are Online Employers
  Fair and Honest?} \date{}

\ifthenelse{\equal{\draft}{true}}{
\maketitle
}{  
\numberofauthors{1}

\author{
\alignauthor
John J. Horton\\
       \affaddr{Harvard University}\\
       \affaddr{383 Pforzheimer Mail Center}\\
       \affaddr{56 Linnaean Street}\\
       \affaddr{Cambridge, MA 02138}\\
       \email{john.joseph.horton@gmail.com}
}
\maketitle
}

\begin{abstract} 
Online labor markets give people in poor countries direct access to
buyers in rich countries. Economic theory and empirical evidence
strongly suggest that this kind of access improves human
welfare. However, critics claim that abuses are endemic in these
markets and that employers exploit unprotected, vulnerable workers. I
investigate part of this claim using a randomized, paired survey in
which I ask workers in an online labor market (Amazon Mechanical Turk)
how they perceive online employers and employers in their host country
in terms of honesty and fairness. I find that, on average, workers
perceive the collection of online employers as slightly fairer and
more honest than offline employers, though the effect is not
significant. Views are more polarized in the online employer case,
with more respondents having very positive views of the online
collection of employers.
\end{abstract}

\ifthenelse{\equal{\draft}{true}}{}{
}

\setkeys{Gin}{width=0.45\textwidth}

\section{Introduction} \label{sec:major}
Work conducted over the Internet by workers participating in online
labor markets has begun to attract mainstream attention and much of
this attention has been negative. Most of the criticism targets
low-skilled, low-paying piece-work sites like Amazon's Mechanical Turk
(AMT), though other sites like oDesk and Elance are also drawing
scrutiny. Harvard Law Professor and co-founder of the Berkman Center
for Internet \& Society Jonathan Zittrain recently published an
article \emph{Newsweek} titled
\href{http://www.newsweek.com/id/225629}{``Work the New Digital
  Sweatshops.''} At a recent conference on
\href{http://digitallabor.org/}{digital labor} at the New School, the
words ``expropriation'' or ``exploitation'' were on nearly every page
of the program. Critics worry that buyers are circumventing labor laws
and exploiting workers.  Aside from low pay, critics argue that
workers do not know the (potentially unethical) purpose of their work
and have no ability to organize or appeal the decisions of capricious
employers \cite{zittrain2008ubiquitous, irani2009}. They also worry
that much of the work is of dubious social value, with many buyers
using workers to generate spam and write
\href{http://www.crunchgear.com/2009/01/17/belkin-paying-65-cents-for-good-reviews-on-newegg-and-amazon/}{bogus}
product reviews.

Despite these perceived downsides of online labor markets, they have a
tremendous and potentially transformative upside, which is that the
markets give people in poor countries access to buyers in rich
countries. If this form of increased virtual labor mobility has
effects similar to those of increased real labor mobility, then the
emergence of online labor markets should be lauded and supported; the
welfare gains from liberalizing restrictions on labor mobility are
truly enormous.  Clemens et al. \cite{clemens2008place} consider the
effect relocation to the US would have on the \emph{real} wages of
workers from different countries. For the median country (Bolivia),
wages would increase by a factor of 2.7, and for the highest country
(Nigeria), wages would increase by a factor of 8.4. Even with current
strict limits on migration, the
\href{http://web.worldbank.org/WBSITE/EXTERNAL/NEWS/0,,contentMDK:20648762~menuPK:34480~pagePK:64257043~piPK:437376~theSitePK:4607,00.html}{World
  Bank} estimates that in 2008 remittances to developing countries
were over \$305 billion, which exceeds both private capital flows and
official development aid.

The comparative advantages of the world's poor are that they (either
individually or collectively through political institutions) are
willing to accept environmental degradation, dangerous working
conditions and very low pay. In light of this unpleasant truth, the
relative virtues of digital work are obvious: it poses no physical
danger to workers, has virtually no environmental impact and it does
not require robust host country institutions or local entrepreneurial
talent. Workers can set their own hours and are not exposed to the
elements, dangerous working conditions, the vagaries of agriculture or
tyrannical bosses.  Unlike labor market access gained through physical
migration, workers do not have to live apart from their families or
dissipate their earnings by paying developed country prices for
shelter, food and clothing.

We have discussed both the perceived costs and the potential benefits
of online labor markets; good public policy will require some attempt
to quantify the trade-offs under different policy scenarios. Consider
a proposal to require Amazon to verify that each new task is not being
used to generate spam. While this may have the intended effect of
reducing spam, compliance costs might exceed the per-transaction
profits, thus pricing all work out of the market---or it might
not, leading to an overall gain in welfare. Fortunately, we do not
need to limit ourselves to speculation and conjecture: it is easy to
test hypotheses about online phenomena by running experiments.

\section{The Experiment}
In AMT, the decision whether to pay a worker is left wholly up to the
buyer; nothing stops a buyer from expropriating the product of
workers. Given that the rules tilt heavily in favor of buyers, one
might presume that employers in AMT would regularly cheat their
workers. If this is true, then AMT workers would have dim views of the
honesty and fairness of online employers vis-a-vis offline employers
who operate in the shadow of formal sanctions for unethical
behavior. To test this proposition, I conducted a simple experiment in
which subjects recruited from AMT were asked to answer one of the
following questions:

\begin{itemize} 
\item What percentage (between 0 and 100) of Employers in your home
country would you estimate treat workers honestly and fairly?

\item What percentage (between 0 and 100) of Mechanical Turk
  Requesters would you estimate treat workers honestly and fairly?
\end{itemize} 

Subjects were randomly assigned to a question and only saw that
particular question. In other words, subjects asked about home country
bosses were not asked about requesters and vice versa.  I launched the
experiment on December 23rd and left it open for 7 days. In total, 200
subjects participated. I paid 12 cents per response, which gave an
hourly average wage of \$5.68 and my total expenditure was
\$26.40. Consistent with Amazon's guidance on what constitutes a
``good'' feedback score, the HIT was limited to AMT workers with a
95\% approval rating. Exactly 200 subjects completed the HIT, but only
192 responses were usable.\footnote{The author's website contains the
  raw data from AMT, the code that cleans the data and all code used
  for the analysis and plots. The main cause was workers offering a
  range rather than a point estimate.}  When asked for their home
country, 111 reported being from the US,
58 reported being from India and
23 reported being from some other
country.

\subsection{Results} 
\begin{figure} 
  \begin{center} 
\includegraphics{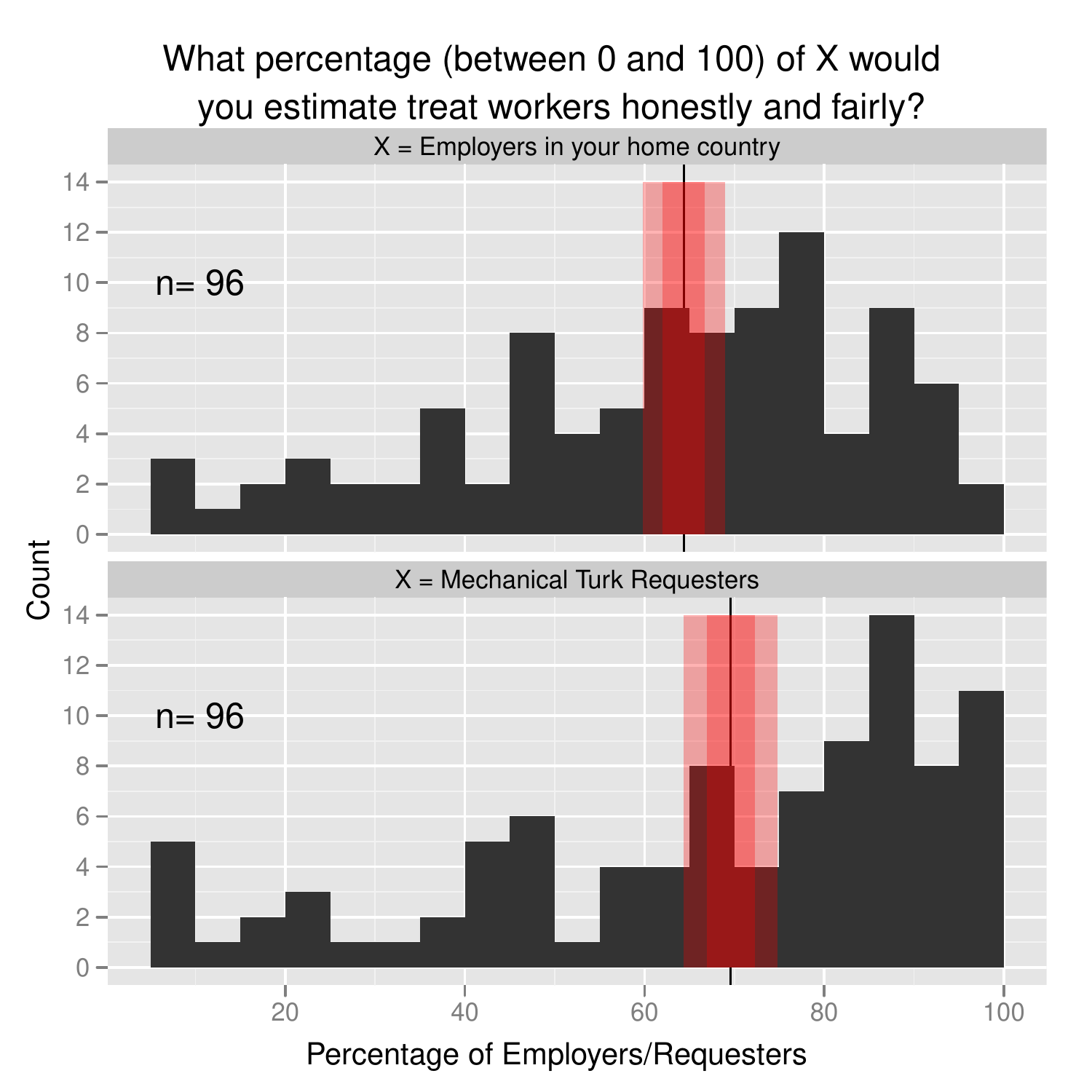}
\caption{Worker attitudes about online and offline employers  \label{fig:hist}}
\end{center} 
\end{figure}

A histogram of worker responses by question type, with a bin width of
5, is shown in Figure \ref{fig:hist}. In each panel, the mean response
is indicated with a vertical line, and 90\% and 95\% confidence
intervals are shown with red shaded bands around the mean. We can see
that (1) means are quite similar and (2) that response in the online
case (bottom panel) appear more polarized, with a greater number of
subjects in the online case having very positive views of the
requesters.

We can compare mean responses in the two groups using a linear
regression of the reported percentage of honest and fair employers,
$perc_i$, on an indicator for whether the subject was asked about
online employers, $AMT_i=1$ or offline, host country
employers, $AMT_i=0$. The fitted regression line, with robust
standard errors, is:
\[ 
  perc_{i} = \underbrace{5.208}_{3.58} \cdot AMT_i +
  \underbrace{64.375}_{2.36}\] with $R^2= 0.011$ and
  $N=192$. We can see that the mean percentage was a little more
  than 64\% and for online employers slightly more than 69\%. The
  difference is not statistically significant.

Confirming the graphical evidence that more workers have very positive
views of online employers compared to offline employers, a regression
of an indicator for whether the subject perceived that more than 80\%
of employers were honest and fair yields a positive and highly
significant coefficient on the group assignment indicator, $AMT_i$:
\[ 
  1\{perc_{i} > 80\} = \underbrace{0.219}_{0.07} \cdot AMT_i +
 \underbrace{0.219}_{0.04}\] with $R^2= 0.05426$ and
 $N=192$. 

There are several caveats to the findings. Experimenter effects could
matter: subjects might exaggerate how honest and fair they find AMT
employers, because this question was asked by an AMT employer. An
unavoidable limitation is that subjects are not a random sample of AMT
workers---perhaps the ones who have bad experiences quit. The 95\%
cutoff might preclude the participation of a large number of
disgruntled workers, though this seems unlikely; in past experiments,
I have found very little difference in uptake under different cutoffs,
suggesting most workers have high scores.

\section{Discussion}
The critique of online markets goes beyond the perceived fairness of
employers. Furthermore, worker perceptions of fairness are not
measures of \emph{actual} fairness. That being said, the experiment
offers evidence that AMT workers view their chances of being treated
fairly online as being as good or better than what they can obtain
offline. Contrary to our prior expectations, rampant exploitation is a
mis-characterization.

Future research should investigate other claims related to online
markets: how prevalent are dubious tasks? Do workers get repetitive
stress injuries, as is often suggested? Do workers feel they are
gaining skills? Answers to these questions could help clarify the
trade-offs inherent in different policy proposals. Online work is
currently a small phenomena compared to the global trade in services,
but it will become far larger and will eventually attract more
policy-oriented attention. Given the welfare consequences of online
work, it would be a tragedy if supposition and conjecture about easily
and cheaply answerable empirical questions determined our digital
policy.

\section{Acknowledgments} 
John Horton thanks the Berkman Center for Internet and Society and the
NSF-IGERT Multidisciplinary Program in Inequality \& Social Policy for
generous financial support. Thanks to Aaron Shaw, Roberta Horton and
Carolyn Yerkes for helpful comments and suggestions. All plots were
made using the open source R package \texttt{ggplot2}, developed by
Hadley Wickham \cite{wickham2008ggplot2}.

\bibliographystyle{abbrv}
\bibliography{masterDrop.bib}

\end{document}